\documentclass{article}
\usepackage{amsmath}

\input{tcilatex}
\begin{document}

\title{Exchange of quantum states between coupled oscillators\ }
\author{D. Portes Jr. and H. Rodrigues \\
%EndAName
Centro Federal de Educa\c{c}\~{a}o Tecnol\'{o}gica do Rio de Janeiro\\
Departamento de Educa\c{c}\~{a}o Superior-DEPES. \and S. B. Duarte\thanks{%
corresponding author, e-mail : sbd@cbpf.br} \\
%EndAName
Centro Brasileiro de Pesquisas F\'{\i}sicas /CNPq, \ \\
Rua Dr. Xavier Sigaud 150, CEP 22.290-180,\\
Rio de Janeiro (RJ), Brazil. \and B. Baseia \\
%EndAName
Instituto de F\'{\i}sica, Universidade Federal de Goi\'{a}s\\
PO-Box-131, 74.001-970, Goiania(GO), Brazil.}
\maketitle

\begin{abstract}
\emph{Exchange of quantum states between two interacting harmonic oscillator
along their evolution time is discussed. It is analyzed the conditions for
such exchange starting from a generic initial state and demonstrating that
the effect occurs exactly only for the particular states }$C_{0}\ |0>+C_{N}\
|N>,$\emph{\ which includes the interesting qubits components }$|0\rangle
,|1\rangle $.\emph{\ It is also determined the relation between the coupling
constant and characteristic frequencies of the oscillators to have the
complete exchange. }
\end{abstract}

\section{\ Introduction}

The engineering of quantum states of light fields and oscillators became an
interesting topic in the last years, due to its applications in : (i)
fundamentals of quantum mechanics (preparation of Schrodinger-cat states 
\cite{Yurke}, their superposition \cite{Davidovich} and measurement of their
decoherence \cite{Brune}, etc.); (ii) determination of certain properties of
a system (phase distribution P($\theta $) \cite{Pegg}, Wigner \cite{Wigner}
and Husimi \cite{Moussa} functions, etc.); (iii) proposals for practical
applications (quantum lithography \cite{Bjork}, quantum communication \cite%
{Braustein} - e.g., via hole-burning in Fock space \cite{Baseia} - quantum
teleportation \cite{Julsgaard}, etc). However, a difficult situation appears
when one wants to prepare a state of a system offering hard access \cite%
{Dietrich}. In this case the difficulty may be circumvented by coupling the
system having hard access to a second system offering easy access, in which
a desired state is prepared with subsequent transfer to the first one. The
success of this operation depends on the model-Hamiltonian and on the
initial state describing the whole system.

Although the problem of two interacting harmonic oscillators has been
exhaustively studied \ in the literature, the discussion about exchange of
nonclassical states between them\textbf{\ is} scarce. The coupled quantum
oscillation problem was considered earlier in \cite%
{Oliveira6,Oliveira7,Oliveira17}, where the authors of those papers were
interested only in the energy of the system.\ Later on, in Ref \cite%
{Oliveira} a full exchange between quantum two-mode harmonic oscillators
was\ presented, however the issue was only concerned with the particular
transfer of coherent states.\textbf{\ }In Ref. \cite{Rodrigues} we have
studied the transfer of certain properties (statistics and squeezing) and in
Ref. \cite{Portes} we have studied the transfer of \textbf{the most}
relevant part of the state of a sub-system to another, through the
simultaneous transfer of the number and phase distributions, $P_{n}$ and $%
P(\theta )\footnote{%
Since the number and phase are canonically conjugate operators they are
complementary, in the sense that simultaneous transfer of number and phase
distributions, $P_{n}$ and $P(\theta )$, concerns the transfer of the major
part of the state describing a system.}$ \cite{Portes}; the solutions were
found numerically since the models were not exactly soluble.

In the present work we employ a distinct model-Hamiltonian, allowing us to
treat the problem analytically permitting us to analyze the transfer of
generic states. We show in which way one can get exact exchange of the
states between two interacting sub-systems. \textit{Exchange of states}
means \textit{simultaneous} \textit{transfer of states in two opposite
directions}; so, it is more significant than the transfer of states in one
direction as studied in \cite{Portes}. In the present case the transfer of a
state from the \textquotedblleft \textit{easy-oscillator\textquotedblright }
to the \textquotedblleft \textit{hard-oscillator}\textquotedblright\ is
observed by simply monitoring the state of the easy-oscillator during the
time evolution of the whole system. For brevity, hereafter the easy- and the
hard-oscillator will be referred to as \textbf{O}$_{1}$ and \textbf{O}$_{2}$%
, respectively.

The Sect. II introduces the model-Hamiltonian allowing us to obtain the
evolution operator for this coupled system. In the Sect. III we consider
different types of initial states describing the entire system to study the
mentioned effect between the \textbf{O}$_{1}$ and the \textbf{O}$_{2}$ (
Sub-Sects. \textbf{(A)}, \textbf{(B)},and \textbf{(C) }), including
superpositions of states representing the qubits $|0\rangle $ and $|1\rangle 
$. The Sect. IV contains the comments and conclusion.

\section{\protect\bigskip Model-Hamiltonian: evolution operator}

We start from the Hamiltonian 
\begin{equation}
\mathbf{H/}\hbar =\omega _{1}\mathbf{a}_{1}^{+}\mathbf{a}_{1}+\omega _{2}%
\mathbf{a}_{2}^{+}\mathbf{a}_{2}+\lambda \,\left( \mathbf{a}_{1}^{+}\mathbf{a%
}_{2}+\mathbf{a}_{1}\mathbf{a}_{2}^{+}\right) \;,  \label{1}
\end{equation}%
where $\mathbf{a}_{i}^{+}$($\mathbf{a}_{i})$ stands for the raising
(lowering) operator of the $i-th$ oscillator, \ $i=1,2$; \ $\omega _{i}$ and 
$\lambda $ are real parameters standing for the i-th oscillator frequency
and coupling constant, respectively. The equations of motion for the
operators $a_{1}(t)$ and $a_{2}(t)$\ can be solved analytically, 
\begin{eqnarray}
\mathbf{a}_{1}(t) &=&\left( c^{2}e^{-i\omega _{1}^{\prime
}t}+s^{2}e^{-i\omega _{2}^{\prime }}t\right) \mathbf{a}_{1}(0)+cs\left(
e^{-i\omega _{1}^{\prime }t}-e^{-i\omega _{2}^{\prime }}t\right) \mathbf{a}%
_{2}(0),  \label{2} \\
\mathbf{a}_{2}(t) &=&\left( c^{2}e^{-i\omega _{2}^{\prime
}t}+s^{2}e^{-i\omega _{1}^{\prime }}t\right) \mathbf{a}_{2}(0)+cs\left(
e^{-i\omega _{1}^{\prime }t}-e^{-i\omega _{2}^{\prime }}t\right) \mathbf{a}%
_{1}(0),  \notag
\end{eqnarray}%
where,%
\begin{eqnarray}
\omega _{1}^{\prime } &=&\omega _{1}+\,\lambda \frac{s}{c}~,  \label{3} \\
\omega _{2}^{\prime } &=&\omega _{2}-\lambda \,\frac{s}{c}\;.  \notag
\end{eqnarray}%
and%
\begin{eqnarray}
s &=&\left( \frac{1}{2}-\frac{x}{2\sqrt{x^{2}+1}}\,\right) ^{1/2}  \label{4}
\\
c &=&\left( \frac{1}{2}+\frac{x}{2\sqrt{x^{2}+1}}\,\right) ^{1/2}  \notag
\end{eqnarray}%
with%
\begin{equation}
x=\frac{\omega _{1}-\omega _{2}}{2\lambda }\ .  \label{5}
\end{equation}%
The parameter $s$\ and $c$\ satisfy the condition $c^{2}+s^{2}=1,$\ they
define the auxiliary operators%
\begin{eqnarray}
\mathbf{a}_{1}^{\prime } &=&c\,\mathbf{a}_{1}+s\,\mathbf{a}_{2}~\text{,}
\label{6} \\
\mathbf{a}_{2}^{\prime } &=&-s\,\mathbf{a}_{1}+c\,\mathbf{a}_{2}~,\;  \notag
\end{eqnarray}%
which\emph{\ }decouple the above Hamiltonian. $\;$The following relations
also hold: 
\begin{eqnarray}
\omega _{1}^{\prime }+\omega _{2}^{\prime } &=&\omega _{1}+\omega _{2}\text{
,}  \label{7} \\
\omega _{1}^{\prime }-\omega _{2}^{\prime } &=&\frac{\lambda }{cs}\text{ .} 
\notag
\end{eqnarray}

It is convenient for our purposes to find the time dependent state vector or
density operator in the Schrodinger picture. One formal prescription is to
work with Wigner representation of the state and obtain the time-dependent
density operator from the Wigner function\cite{Mollow}, for which the time
evolution is easily obtained. However,\ it is a hard task to restore
analytical or numerical values for the density matrix $\rho (t)$\ in the
Fock basis from the time dependent Wigner function. To\ overcome this
difficult we will show that\ for the\textbf{\ }Hamiltonian given by Eq.(\ref%
{1}) there is an analytical expression for the evolution operator $U(t)$,
which defines the solution of the Schrodinger equation, allo\textbf{wing us }%
to get directly the matrix\emph{\ }$\rho (t)$\ in the Fock basis. This kind
of approach was already used in Ref \cite{Lee}, but only treating the system
in the resonant case ($\omega _{1}=\omega _{2}$). \textbf{I}n \cite{Lee}\
the author studied the transfer of state starting from the particular one
photon state. Our results permit one to obtain an analytical expression for
the matrix element $U(t),$\ for the Hamiltonian (\ref{1}) not restricted to
the resonant case and permitting easy application to a generic initial
state. Consequently, the\textbf{\ }problem of transfer of states can be more
comfortably discussed using\ the present results.

To obtain the \textbf{operator} $U(t)$, we define the\textbf{\ (auxiliary})
unitary operator $\mathbf{U}_{s}(t)$\ \textbf{\ }which is associated to a
rotation and decouples the Hamiltonian,\emph{\ }%
\begin{equation}
\mathbf{U}_{s}^{-1}\,\mathbf{a}_{i}\mathbf{\,U}_{s}=a_{i}^{\prime }\ .
\label{8}
\end{equation}

We have, 
\begin{equation}
\mathbf{U}_{s}^{-1}=\mathbf{U}_{-s}\;,  \label{9}
\end{equation}%
in view of \ the reverse transformation

\begin{eqnarray}
\mathbf{a}_{1} &=&c\,\mathbf{a}_{1}^{\prime }-s\,\mathbf{a}_{2}^{\prime }%
\text{ ,}  \label{10} \\
\mathbf{a}_{2} &=&s\,\mathbf{a}_{1}^{\prime }+c\,\mathbf{a}_{2}^{\prime }%
\text{ .}  \notag
\end{eqnarray}

We denote \{$|n_{1},n_{2}\rangle _{0}$\} as representing the Fock$^{\prime }$%
s basis, eigenvectors of the (old) number operator $\mathbf{N}_{i}=\mathbf{a}%
_{i}^{+}\mathbf{a}_{i}$ \textbf{, }whereas $\{|n_{1},n_{2}\rangle _{s}\}$ is
the same for the (new) number operator $\mathbf{N}_{i}(s)=\mathbf{a}%
_{i}^{\prime +}\mathbf{a}_{i}^{\prime }.$ We have, 
\begin{eqnarray}
\mathbf{U}_{s}|n_{1},n_{2}\rangle _{s} &=&|n_{1},n_{2}\rangle _{0},
\label{11} \\
|n_{1},n_{2}\rangle _{s} &=&\mathbf{U}_{-s}|n_{1},n_{2}\rangle _{0}.  \notag
\end{eqnarray}%
If we represent $\mathbf{U}_{s}\,$\ in the Fock$^{\prime }$s basis $%
\{|n_{1},n_{2}\rangle _{0}\},$ we obtain 
\begin{eqnarray}
\left( \mathbf{U}_{s}\right) _{m_{1},\text{ }m_{2}}^{n_{1},\text{ }n_{2}}
&=&\,_{0}\langle n_{1},n_{2}|\mathbf{U}_{s}|m_{1},m_{2}\rangle _{0}
\label{12} \\
&=&\,_{s}\langle n_{1},n_{2}|m_{1},m_{2}\rangle _{0}.  \notag
\end{eqnarray}%
Next, to reconstruct the operator $\mathbf{U}_{s}$ in the Fock's basis, we
start from

\begin{equation}
_{s}\langle n_{1},n_{2}|\,\mathbf{a}_{1}^{\prime }|m_{1},m_{2}\rangle
_{0}=\,_{s}\langle n_{1},n_{2}|\left( c\,\mathbf{a}_{1}+s\,\mathbf{a}%
_{2}\right) |m_{1},m_{2}\rangle _{0},  \label{13}
\end{equation}%
Since\ the operators\ $\mathbf{a}_{i\text{ }}^{\prime }$\ act on\ the basis $%
\{|n_{1},n_{2}\rangle _{s}\}$\ whereas the $\mathbf{a}_{i}$\ act on the\
basis $\{|n_{1},n_{2}\rangle _{0}\},$\ we get 
\begin{equation}
\sqrt{n_{1}+1}_{s}\langle n_{1}+1,n_{2}|m_{1},m_{2}\rangle _{0}=\,c\sqrt{%
m_{1}}\,\,_{s}\langle n_{1},n_{2}|m_{1}-1,m_{2}\rangle _{0}  \label{14}
\end{equation}%
\begin{equation*}
\ \ \ \ \ \ \ \ \ \ \ \ \ \ \ \ \ \ \ \ \ \ \ \ \ \ \ \ \ \ \ \ \ \ \ \ \ \
\ \ \ \ \ \ \ +s\sqrt{m_{2}}\,\,_{s}\langle n_{1},n_{2}|m_{1},m_{2}-1\rangle
_{0},
\end{equation*}%
which,\textbf{\ after} using the Eq.(\ref{12}), leads to

\begin{equation}
\left( \mathbf{U}_{s}\right) _{m_{1},\text{ }m_{2}}^{n_{1},\text{ }n_{2}}=c%
\sqrt{\frac{m_{1}}{n_{1}}}\left( \mathbf{U}_{s}\right)
_{m_{1}-1,m_{2}}^{n_{1}-1,n_{2}}+s\sqrt{\frac{m_{2}}{n_{1}}}\left( \mathbf{U}%
_{s}\right) _{m_{1},m_{2}-1}^{n_{1}-1,n_{2}}\;,  \label{15}
\end{equation}%
and similarly, repeating the procedure for the operator $\mathbf{a}%
_{2}^{\prime }$, we find 
\begin{equation}
\left( \mathbf{U}_{s}\right) _{m_{1},\text{ }m_{2}}^{n_{1},\text{ }n_{2}}=-s%
\sqrt{\frac{m_{1}}{n_{2}}}\left( \mathbf{U}_{s}\right)
_{m_{1}-1,m_{2}}^{n_{1},n_{2}-1}+c\sqrt{\frac{m_{2}}{n_{2}}}\left( \mathbf{U}%
_{s}\right) _{m_{1},m_{2}-1}^{n_{1},n_{2}-1}\;.  \label{16}
\end{equation}

Using the Eqs. (\ref{15}), (\ref{16}) plus the unitary condition $%
U_{s}^{\dagger }U_{s}=U_{s}U_{s}^{\dagger }=1$\ \ we obtain, after a lengthy
calculation, the\ expression 
\begin{eqnarray}
\left( \mathbf{U}_{s}\right) _{m_{1},\text{ }m_{2}}^{n_{1},\text{ }n_{2}}
&=&\delta _{n_{1}+n_{2},\text{ }m_{1}+m_{2}}\sqrt{\frac{n_{1}!n_{2}!}{%
m_{1}!m_{2}!}}\left( -1\right) ^{n_{2}}c^{m_{1}-n_{2}}\,s^{m_{2}+n_{2}}\ \ 
\label{17} \\
&&\times \sum_{k=\max (0,m_{2}-n_{1})}^{\min (n_{2},m_{2})}\left( -1\right)
^{-k}\left( \frac{s}{c}\right) ^{-2k}\,\dbinom{m_{1}}{n_{2}-k}\dbinom{m_{2}}{%
k}\text{ ,}  \notag
\end{eqnarray}%
\newline
and 
\begin{equation}
\left( \mathbf{U}_{-s}\right) _{m_{1},\text{ }m_{2}}^{n_{1},\text{ }%
n_{2}}=\left( -1\right) ^{m_{2}-n_{2}}\left( \mathbf{U}_{s}\right) _{m_{1},%
\text{ }m_{2}}^{n_{1},\text{ }n_{2}}\;.  \label{18}
\end{equation}%
The time evolution operator $U(t)$\ may be written in the basis $%
\{|n_{1},n_{2}\rangle _{s}\}$\ as

\begin{equation}
\mathbf{U(t)}=\sum_{k_{1},k_{2}}|k_{1},k_{2}\rangle _{s}~e^{-i(k_{1}\omega
_{1}^{\prime }+\,k_{2}\omega _{2}^{\prime })t}~_{s}\langle k_{1},k_{2}|\text{
},  \label{19}
\end{equation}%
for $H$\ is diagonal in this basis. Finally from the Eqs.(\ref{12}) and (\ref%
{19}) we obtain the expression

\begin{equation}
\mathbf{U(t)}_{m_{1},\text{ }m_{2}}^{n_{1},\text{ }n_{2}}=%
\sum_{k_{1},k_{2}}e^{-i(k_{1}\omega _{1}^{\prime }+\,k_{2}\omega
_{2}^{\prime })t}\left( \mathbf{U}_{-s}\right) _{k_{1},\text{ }k_{2}}^{n_{1},%
\text{ }n_{2}}\left( \mathbf{U}_{-s}\right) _{k_{1},\text{ }k_{2}}^{m_{1},%
\text{ }m_{2}}\;,  \label{20}
\end{equation}%
restricted to $n_{1}+n_{2}=k_{1}+k_{2}=m_{1}+m_{2}$\ , whereas $U_{m_{1},%
\text{ }m_{2}}^{n_{1},\text{ }n_{2}}=0$\ otherwise.

The evolution operator obtained in Eq.(\ref{20}) allows us to study the time
evolution of the whole state describing our bipartite system composed by
coupled oscillators, represented by the Hamiltonian in the Eq.(\ref{1}). In
the next section we will study the exchange of states between these
oscillators \ and, as a natural assumption, we will suppose the \textbf{O}$%
_{2}$ initially in its ground state$\,|0\rangle $. The \textbf{O}$_{1}$ is
assumed to be{\LARGE \ }previously prepared in various initial states,
firstly starting from an{\LARGE \ }arbitrary state $|\phi \rangle $.

\section{Exchange of generic state}

Let us consider that the whole (bipartite) system is initially in the state 
\begin{equation}
|\Psi (0)\rangle \,=\,|\phi \rangle \otimes |0\rangle \,,  \label{21}
\end{equation}%
whose components in the Fock's \ basis are given by, 
\begin{equation}
|\Psi (0)\rangle \,=\sum_{n}C^{n,\text{ }0}(0)|n,0\rangle \,,  \label{22}
\end{equation}%
since $C^{n_{1},\text{ }n_{2}}(0)=0$ for \thinspace $n_{2}\neq 0.\,$In the
Schrodinger representation, the\ coefficients $C^{n_{1},n_{2}}(t)$ are
obtained from $C^{n_{1},n_{2}}(t)=\,\langle n_{1},n_{2}|\mathbf{U(t)}|\Psi
(0)\rangle ,$ which, using Eq. (\ref{22}) and the constraint $n_{1}+n_{2}=n$%
, results in the form 
\begin{equation}
C^{n_{1},n_{2}}(t)=\,C^{n_{1}+n_{2},0}(0)\,\mathbf{U(t)}%
_{n_{1}+n_{2},0}^{n_{1},n_{2}}\;.  \label{23}
\end{equation}%
In particular, we have that 
\begin{equation}
C^{n,0}(t)=\,C^{n,0}(0)\,\mathbf{U(t)}_{n,0}^{n,0}\;,  \label{24}
\end{equation}%
and 
\begin{equation}
C^{0,n}(t)=\,C^{n,0}(0)\,\mathbf{U(t)}_{n,0}^{0,n}\;.  \label{25}
\end{equation}%
The exchange of states between the oscillators will occur after an instant $%
\tau ,$when $C^{0,n}(\tau )=C^{n,0}(0)$\thinspace and 
\begin{equation}
|\Psi (\tau )\rangle \,=\,\sum_{n}C^{0,\text{ }n}(\tau )|0,n\rangle \,,
\label{26}
\end{equation}%
or, $\ |\Psi (\tau )\rangle \,=\,|0\rangle \otimes |\phi \rangle .$ This
shows that exchange of states allows us to verify the transfer of states to
the \textbf{O}$_{2}$ by monitoring the time evolution of the \textbf{O}$_{1}$%
.

From the Eqs. (\ref{17}) and (\ref{18}) we have,

\begin{equation}
\left( \mathbf{U}_{s}\right) _{n-l,l}^{n,0}=\sqrt{\frac{n!}{(n-l)!l!}}%
c^{n-l}\,s^{l}\text{ ,}  \label{29}
\end{equation}%
and 
\begin{equation}
\left( \mathbf{U}_{s}\right) _{n-l,l}^{0,n}=\sqrt{\frac{n!}{(n-l)!l!}}%
\,\left( -1\right) ^{n-l}c^{l}\,s^{n-l}\;.  \label{30}
\end{equation}%
The substitution of the Eqs. (\ref{29}) and (\ref{30}) in the Eq. (\ref{20})
results 
\begin{equation}
\,\mathbf{U(t)}_{n,0}^{0,n}=\left( -1\right) ^{n}\sum_{l=0}^{n}\frac{n!}{%
(n-l)!l!}\,\left( -1\right) ^{n-l}c^{n}s^{n}e^{-i\,(n-l)~\omega _{1}^{\prime
}\,\,t}\,e^{-i\,l~\omega _{2}^{\prime }\,\,t}\;.  \label{31}
\end{equation}%
where we recognize the Newton's binomial expression, 
\begin{equation}
\mathbf{U(t)}_{n,0}^{0,n}=\left( -1\right) ^{n}c^{n}s^{n}\,\left(
e^{-i\,\omega _{2}^{\prime }\,\,t}-e^{-i\,\omega _{1}^{\prime }\,\,t}\right)
^{n}  \label{32}
\end{equation}%
or, replacing the auxiliary parameters $\omega _{1}^{\prime }$, $\omega
_{2}^{\prime }$ by $\omega _{1},$ $\omega _{2}$ and $\lambda $ (cf. Eq. (\ref%
{7})), 
\begin{equation}
\mathbf{U(t)}_{n,0}^{0,n}=e^{-i\frac{\omega _{1}+\omega _{2}}{2}%
\,n\,t}\left( -2\,i\,s\,c\sin (\frac{\lambda }{2\,cs}\,t)\right) ^{n}\;,
\label{33}
\end{equation}%
\emph{\ }and, consequently, 
\begin{equation}
C^{0,n}(t)=C^{n,0}(0)\,e^{-i\frac{\omega _{1}+\omega _{2}}{2}\,n\,t}\left(
-2\,i\,s\,c\sin (\frac{\lambda }{2\,cs}\,t)\right) ^{n}\text{ .}  \label{34}
\end{equation}%
In a similar way we get, 
\begin{equation}
C^{n,0}(t)=C^{n,0}(0)\,e^{-i\,\frac{\omega _{1}+\omega _{2}}{2}\,n\,t}\left(
c^{2}e^{-i\,\frac{\lambda }{2cs}\,\,t}+s^{2}e^{i\,\frac{\lambda }{2cs}%
\,\,t}\right) ^{n}\;.  \label{35}
\end{equation}%
\qquad

From Eq. (\ref{34}) we see that\emph{\ }a partial exchange of states will
occur when $\lambda t/sc=\left( 2k+1\right) \pi ,\,$\ i.e., in the time
intervals $\tau _{k}=\left( sc/\lambda \right) $\ $\left( 2k+1\right) \pi .$
The effect attains the highest efficiency when the product $sc$ is maximum,
i.e., when $s=c=1/\sqrt{2}$ and $\tau _{k}=$\ $\left( k+1/2\right) \pi
/\lambda $. According to the Eq. (\ref{4}) this implies $x=0$ and the
resonance condition $\omega _{1}=\omega _{2}=\omega $ (cf. Eq. (\ref{5})), 
\begin{equation}
C^{0,n}(\tau _{k})=(-i)^{n}\,C^{n,0}(0)\,e^{-i\,\omega \,n\,\tau _{k}}.
\label{36}
\end{equation}%
However, we note that even at resonance we obtain no exchange of states, due
to the presence of the phase factor $\exp \left[ -i\,\left( \omega \tau _{k}+%
\frac{\pi }{2}\right) \,n\right] $ affecting the coefficients of the state
describing both oscillators in the Fock's representation. In this general
case we obtain $\left\vert C^{0,n}(\tau _{k})\right\vert =\left\vert
C^{n,0}(0)\right\vert $,\textbf{\ }which means exchange of statistics
between the two oscillators. This can also be seen comparing both reduced
density matrix, $\rho _{m_{1},\text{ }m_{2}}^{(2)}(\tau _{k})$ \ and $\rho
_{m_{1},\text{ }m_{2}}^{(1)}(0)$, in the Fock's representation,

\begin{equation}
\rho _{m_{1},\text{ }m_{2}}^{(2)}(\tau _{k})=e^{-i\,\left( \omega \tau _{k}+%
\frac{\pi }{2}\right) \,(m_{1}-m_{2})}\text{ }\rho _{m_{1},\text{ }%
m_{2}}^{(1)}(0)\text{ },  \label{37}
\end{equation}%
which exhibits the distinction between their off-diagonal elements. As well
known, while the\ state of a system offers its complete description, the
same is not true for the statistics, which contains only partial
informations of the system.

\subsection{The complete exchange of state}

It is shown in the last section that it is not possible to have a complete
exchange of states for a generic initial state because the phases are not
transferred (see Eq.\ref{37}). Here we show that when the state of
oscillator O$_{1}$ is given by the superposition $C_{0}|0\rangle
+\,C_{N}|N\rangle $\ whereas O$_{2\text{ }}$ is in the vacuum state,
complete exchange of states occurs. Note that this state includes in
particular the important case $C_{0}|0\rangle +\,C_{1}|1\rangle $ using the
qubits $|0\rangle $, $|1\rangle $\ having potential applications in quantum
communication \cite{Enk} and in quantum computation \cite{Shor}. It was
shown that this state exhibits squeezed fluctuations \cite{Wodkiewicz}.

Next, let us consider the whole system initially in the superposed state

\begin{equation}
|\Psi (0)\rangle \,=\,C^{0,0}(0)|0,0\rangle +C^{N,0}(0)|N,0\rangle \,.
\label{38}
\end{equation}

In this case we verify perfect exchange of states between the oscillators
for a convenient choice of the parameters involved. Assuming the resonance
condition in the\ Eq.(\ref{34}) we have, for $C^{0,0}(t)=C^{0,0}(0),$

\begin{equation}
C^{0,N}(t)=C^{N,0}(0)\,e^{-i\,\left( \omega \,\,t+\pi /2\right) \,N}\sin
^{N}(\lambda \,t)\;.  \label{39}
\end{equation}%
Partial{\LARGE \ }exchange of states will occur when $t=\tau _{0}=\pi
/(2\lambda ),\,$which results in 
\begin{equation}
C^{0,N}(\tau _{k})=C^{N,0}(0)\,e^{-i\,\pi /2\left( \omega /\lambda +1\right)
\,N}\;,  \label{40}
\end{equation}%
whose meaning\textbf{\ }is the exchange of\textbf{\ }statistics. The
exchange of states becomes complete (exact) when $C^{0,N}(\tau
_{k})=C^{N,0}(0)$, namely, when 
\begin{equation}
\frac{\omega }{\lambda }=\frac{4m-N}{N}\;\text{\ ,}  \label{42}
\end{equation}%
with $m$ integers. Taking $m=1$ \ and $\omega $ in the microwave domain ($%
\omega \sim 10^{9}Hz$) the time spent to transfer the state $%
C_{0}|0>+\;C_{1}|1>$ \ from the \textbf{O}$_{1}$ to the \textbf{O}$_{2}$
results $\tau _{0}=\pi /(2\lambda )$ $\sim $ $10^{-9}$s, since $\lambda
=\omega /3$ (cf. Eq.(\ref{42})){\LARGE ,} which is smaller than the typical
decoherence time for such systems ($\tau _{d}\sim 10^{-3}s)$, as it should.

Note that the previous initial state $C_{0}|0\rangle +\,C_{N}|N\rangle $
describing the O$_{1}$ includes the Fock states $|N\rangle $, obtained from $%
C_{0}=0$ and $C_{N}=1$. In this case exact exchange of states no longer
requires the Eq. (\ref{42}). The reason comes from the phase factor{\LARGE \ 
}appearing in the Eq. (\ref{42}), now becoming a global phase with no
physical relevance. In this case the exchange of states is exact for any
instant $\ t_{k}=\tau _{0}+2\pi k/\lambda .$

\section{Comments and Conclusion}

An analytical procedure applied to a convenient model-Hamiltonian describing
two coupled oscillators allows us to get the exact evolution operator for
the entire system (Sect. II). This approach, through the use of distinct
initial states and parameters (Sub-Sects. \textbf{(A), (B) }of Sect.\textbf{%
\ }III), makes easy the study of exchange of states between such
sub-systems. In all cases we have shown that the fidelity of the process is
maximum when the resonance condition, $\omega _{1}=\omega _{2}$, is
attained. Assuming the \textbf{O}$_{2}$ always in the vacuum state we find,
sub-Section by sub-Section, that: (\textbf{A}) partial exchange of states is
achieved when the initial state of the \textbf{O}$_{1}$ is arbitrary, for
the time intervals $t=\tau _{k}$ $=(k+1/2)\pi /\lambda $; the efficiency of
partial exchange is maximum when the product $sc$ is maximum ($sc=1/2$);
however,\ while the occurrence of exchange of states is partial, exchange of 
\textit{statistics }is obtained exactly, as shown in the Eqs. (\ref{36}), (%
\ref{37}); \textbf{(B)} exact exchange of states occurs when the \textbf{O}$%
_{1}$ starts from the initial superposed state $C_{0}|0\rangle
+\,C_{N}|N\rangle $, in the time intervals $t_{k}=\tau _{0}+2\pi k/\lambda $%
, with the requirement in Eq. (\ref{42}). If the Eq.(\ref{42}) is not
obeyed, exchange of states will occur at the same time intervals, but now
the effect is only partial; Exact exchange of states is also found in the
particular case of (B), setting $C_{0}=0$ and $C_{N}=1$, which means the 
\textbf{O}$_{1}$ starting from a Fock state $|N\rangle $. In this case the
exchange of states occurs exactly at the same time intervals found in (B),
no matter the Eq. (\ref{42}) is obeyed or not.

As final remarks we mention that exchange of states and its efficiency could
be investigated for other model-Hamiltonians and, as explained before, the
effect goes beyond those studied in \cite{Rodrigues} and \cite{Portes}. To
our knowledge, exchange of states in coupled systems and even exchange of
certain properties, are subjects receiving little attention in the
literature \cite{Castro}{\LARGE \ }- with the remarkable exception of
quantum teleportation \cite{Shor}, an effect having a very distinct nature
(requiring the presence of quantum channels and entangled states), which
occurs in the absence of coupling between the two sub-systems. In the
context of teleportation, exchange of states appears with the name "identity
interchange"\ \cite{moussa} and "two-way teleportation" \cite{Vaidman}.

\subsection{Acknowledgements}

The authors thank the CNPq (SBD, BB) and FAPERJ (DPJ) for the partial
supports.

\end{document}